\documentclass[10pt]{iopart}

\usepackage{iopams}  
\usepackage{graphicx}

\begin{document}

\title[Global magnetic cycles in rapidly rotating younger suns]{Global magnetic cycles in rapidly rotating younger suns}

\author{Nicholas J. Nelson$^1$, Benjamin P. Brown$^2$, Matthew K. Browning$^3$, Allan Sacha Brun$^4$, Mark S. Miesch$^5$ and Juri Toomre$^1$}

\address{$^1$ JILA and Department of Astrophysical and Planetary Sciences, University of Colorado, Boulder, CO  80309-0440, USA}
\address{$^2$ Department of Astronomy, University of Wisconsin, 475 Charter St., Madison, WI 53706, USA}
\address{$^3$ Canadian Institute for Theoretical Astrophysics, University of Toronto, Toronto, ON M5S3H8, Canada}
\address{$^4$ DSM/IRFU/SAp, CEA-Saclay, 91191 Gif-sur-Yvette, France}
\address{$^5$ High Altitude Observatory, NCAR, Boulder, CO 80307-3000, USA}
\ead{nnelson@lcd.colorado.edu}
\begin{abstract}
Observations of sun-like stars rotating faster than our current sun tend to 
exhibit increased magnetic activity as well as magnetic cycles spanning 
multiple years. Using global simulations in spherical shells to study the coupling 
of large-scale convection, rotation, and magnetism in a younger sun, we 
have probed effects of rotation on stellar dynamos and the 
nature of magnetic cycles. Major 3-D MHD simulations carried out at three times the 
current solar rotation rate reveal hydromagnetic dynamo action that yields wreaths of strong toroidal magnetic field at low latitudes, often 
with opposite polarity in the two hemispheres.  Our recent 
simulations have explored behavior in systems with considerably lower 
diffusivities, achieved with sub-grid scale models including a 
dynamic Smagorinsky treatment of unresolved turbulence. The lower diffusion promotes the generation of magnetic wreaths that undergo prominent temporal
variations in field strength, exhibiting global magnetic cycles that 
involve polarity reversals. In our least diffusive simulation, we find that magnetic buoyancy coupled with advection by convective giant cells can lead to the rise of coherent loops of magnetic field toward the top of the simulated domain.
\end{abstract}

%\maketitle

\section{Coupling rotation, convection, and magnetism in younger suns}

Global-scale magnetic fields and cycles of magnetic activity in sun-like stars are generated by the interplay of rotation and convection.  At rotation rates greater than that of the current sun, such as when our sun was younger, observations tend to show increased magnetic activity indicating a strong global dynamo may be operating (Pizzolato et al. 2003).  Here we explore large-scale dynamo action in sun-like stars rotating at three times the current solar rate, or $3 \Omega_{\odot}$, with a rotational period of 9.32 days.
%Gobal-scale dynamo action in the sun is believed to arise from the interaction of turbulent convection, rotation, and stratification.  
As shown by helioseismology, the solar interior is in a state of prominent differential rotation in the convection zone (roughly the outer 30\% by radius) whereas the radiative interior is in uniform rotation.  A prominent shear layer, or tachocline, is evident at the interface between the convective and radiative regions.  Motivated by these observations, a number of theoretical models have been proposed for the solar dynamo.  The current paradigms for large-scale solar dynamo action favor a scenario in which the generation sites of toroidal and poloidal fields are spatially separated (e.g., Charbonneau 2005).  Poloidal fields generated by cyclonic turbulence within the bulk of the convection zone, or by breakup of active regions, are pumped downward to the tachocline of rotational shear at its base. The differential rotation there stretches such poloidal fields into strong toroidal structures, which may succumb to magnetic buoyancy instabilities and rise upward to pierce the photosphere as curved structures that form the observed active regions.  Similiar dynamo processes are believed to be active in sun-like stars rotating several times faster than the current sun.  Here we explore a variation to this paradigm by excluding the tachocline and the photosphere from our simulated domain, which extends from $0.72 R_{\odot}$ to $0.96 R_{\odot}$, in order to see if magnetic cycles can be realized in the bulk of the convection zone itself.

\begin{figure}
%\vspace*{-0.9 cm}
\begin{center}
\includegraphics[scale=0.85]{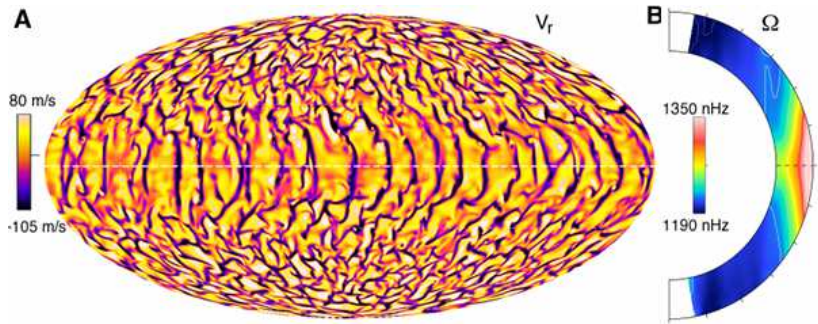}
%\vspace*{-0.3cm}
\caption{ \footnotesize $(A)$~Radial velocity in global Mollweide projection at $0.94 R_{\odot}$ with fast, narrow downflows in dark tones and broad, slow upflows in light tones.  $(B)$~Differential rotation profile, with lines of constant angular velocity $\Omega$ largely along cylinders, as expected for rapidly rotating systems. Some deviation toward conical contours is seen at low latitudes.  Magnetic wreaths tend to form in the regions of strong shear near the equator.}
\end{center}
%\vspace*{-0.9cm}
\end{figure}

%\section{3-D simulations of global stellar dynamos}
Using massively-parallel supercomputers, we solve the nonlinear anelastic MHD equations in rotating 3-D spherical shells using the anelastic spherical harmonic (ASH) code (Brun et al. 2004).  The anelastic approximation filters out fast-moving sound and magneto-acoustic waves, allowing us to follow the decidedly subsonic flows in the solar convection zone with overturning times of days to months.  In large-eddy simulation (LES) such as those using ASH, the effects of small, unresolved scales on larger scales must be parameterized using a turbulence closure model. %Such sub-grid scale (SGS) schemes replace molecular diffusion with turbulent diffusion designed to represent unresolved small-scale mixing.

Previous ASH simulations of convective dynamos in sun-like stars rotating at $3 \Omega_{\odot}$ have yielded large-scale wreaths of strong toroidal magnetic field in the bulk of their convection zones (Brown et al 2010).  These wreaths persist for decades of simulation time, remarkably coexisting with the strongly turbulent flows.  Here we explore the effects of decreased levels of diffusion on these wreaths in two simulations, labeled case B and case S.  Case B uses an eddy viscosity that varies with depth as the square root of the mean density.  Case S uses the dynamic Smagorinsky model of Germano et al. (1991), which is based on the assumption of self-similarity in the inertial range of the velocity spectra.  Case S has 50 times less diffusion on average than case B.  Figure 1a shows the radial velocity field for case S near the top of the convection zone with columnar cells at low latitudes and smaller-scale helical convection at higher latitudes.  Figure 1b shows the differential rotation profile for case S with roughly 20\% (250~nHz) contrast in rotation rate between the equator and poles.  The radial velocity patterns and differential rotation for case B are qualitatively similar to Figure 1.

\begin{figure}
%\vspace*{-0.9 cm}
\begin{center}
\includegraphics[scale=0.96]{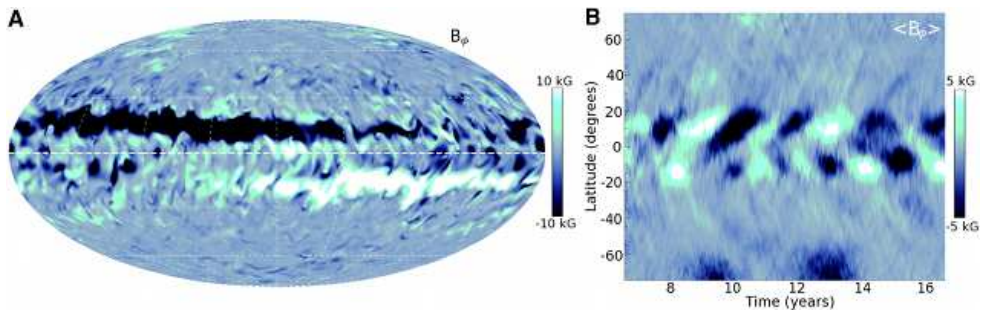}
%\vspace*{-0.3cm}
\caption{ \footnotesize $(A)$~Longitudinal magnetic field $B_\phi$ for case B at $0.84 R_\odot$ in Mollweide projection, showing two strong but patchy magnetic wreaths of opposite polarity with peak field strengths of 38 kG.   $(B)$~Time-latitude plot of $B_\phi$ averaged over longitude $\left< B_{\phi} \right>$ at the same depth over 15 years in case B, with strong negative-polarity wreaths shown in dark tones and strong positive-polarity wreaths shown in light tones, clearly indicating cyclic behavior and reversals in magnetic polarity. }
\end{center}
%\vspace*{-0.9cm}
\end{figure}

\section{Global magnetic cycles and buoyant magnetic loops}

The most remarkable feature of case B is a cyclic variation in the toroidal wreaths of magnetic field.  With significantly less diffusion than the simulation of Brown et al. (2010) that produced persistent wreaths with no reversals, case B creates strong toroidal bands of magnetic field as shown in Figure 2 with peak field strengths of about 38 kG. These wreaths of magnetic field vary strongly with time in both polarity and amplitude.  Figure 2a shows $B_\phi$ in the lower convection zone when there are strong wreaths of opposite polarity in each hemisphere with significant longitudinal variation, which we term patchy wreaths.  If we average over longitude,   Figure 2b shows a time-latitude map of the $\left< B_\phi \right>$ in the lower convection zone.  The simulation clearly goes through reversals in the magnetic polarity of the wreaths in each hemisphere.  At times the hemispheres are out of phase with each other, occasionally yielding wreaths of the same polarity in both hemispheres.  Such behavior might be termed irregular magnetic activity cycles.

\begin{figure}
%\vspace*{-0.9 cm}
\begin{center}
\includegraphics[scale=0.96]{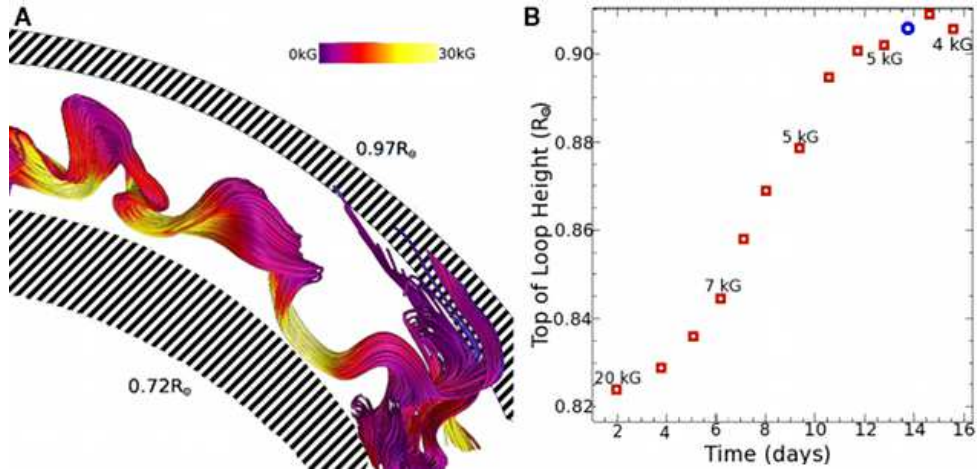}
%\vspace*{-0.3cm}
\caption{ \footnotesize $(A)$~From case S, 3-D volume visualization of magnetic field lines in the core of a wreath-segment with the inner and outer simulation boundaries shown as lined surfaces.  View is looking at low latitudes along the rotation axis. $(B)$~Radial location of the top of a buoyant loop as a function of time.  Magnetic field strength at the top of the loop is indicated at representative times.  Time corresponding to $A$ is indicated by circular plotting symbol at day 13.7.
%While initially the magnetic field is strong enough provide significant magnetic buoyancy, the continued rise is driven at least in part by advection from a convective giant cell. 
}
\end{center}
%\vspace*{-0.9cm}
\end{figure}

%\section{Buoyant magnetic loops}

As we move to even less diffusive simulations, case S shows additional features in the strong toroidal wreaths, most notably buoyant loops of magnetic field.  The wreaths are again patchy in longitude and roughly cyclic in time.  The peak magnetic field strength rises to about 45 kG inside the wreaths.  These strong fields combine with the very low levels of diffusion to allow regions of very strong field to coherently move upward without changing the magnetic topology via reconnection or simply diffusing away the strong fields.  Such magnetic loops rise due to a combination of magnetic buoyancy and advection by convective giant cells that span the layer.  Figure 3a shows a magnetic loop near its maximum size, extending from $0.765$ to $0.908 R_\odot$.  Examination reveals that there is a significant amount of twist present in the loops and that there is a significant deflection poleward as they rise.  The radial location of the top of one buoyant loop as a function of time is shown in Figure 3b.  Initially the buoyancy of the wreath due to evacuation of fluid from magnetic pressure dominates over the advective force of the convective upflows, but within 6 days advection becomes dominant.  After about 10 days the magnetic tension force begins to balance the advection, causing the top of the loop to stall near $0.905 R_\odot$.

%\section{Global dynamos in rapidly rotating suns}

These simulations suggest that stars rotating slightly faster than the current sun may produce dynamos capable of cycles of magnetic activity and buoyant magnetic structures in the bulk of their convective envelopes despite the absence of a tachocline of shear.  This both challenges and informs the interface dynamo paradigm for sun-like stars.  The essential questions are what drives the magnetic reversals in these simulations and what are the conditions necessary to generate buoyant magnetic loops that can survive transit through the convection zone.\\

{ This work is supported by NASA Heliophysics Theory Program grants NNG05G124G and NNX08AI57G and major supercomputing support through NSF TeraGrid resources.  The presentation of this paper in IAU Symposium 273 was aided by NSF grants ATM 0548260 and AST 0968672, and NASA grant 09-LWSTRT09-0039. Browning is supported by the Jeffrey L. Bishop fellowship at CITA.}

\section*{References}
\begin{harvard}
\item[] Brown, B.P., Browning, M.K., Brun, A.S., Miesch, M.S., \& Toomre, J., 2010, ``Persistent magnetic wreaths in a rapidly rotating sun'' {\it Astrophys. J.} {\bf 711} 424
\item[] Brun, A.S., Miesch, M.S., \& Toomre, J., 2004, ``Global-scale turbulent convection and magnetic dynamo action in the solar envelope'', {\it Astrophys. J.} {\bf 614} 1073
\item[] Charbonneau, P., 2005, ``Dynamo models of the solar cycle'', {\it Living Rev. Sol. Phys.} {\bf 2} 2
%\item[] Clune, T.C., Elliot, J.R., Miesch, M.S., Toomre, J., \& Glatzmaier, G.A., 1999, ``Computational aspects of a code to study rotating turbulent convection in spherical shells'', {\it Parallel Comput.} {\bf 25} 4
\item[] Germano, M., Piomelli, U., Moin, P., \& Cabot, W., 1991, ``A dynamic subgrid-scale eddy viscosity model", {\it Phys. Fluids A} {\bf 3} 7
\item[] Pizzolato, N., Maggio, A., Micela, G., Sciortino, S., \& Ventura, R., 2003, ``The stellar activity-rotation relationship revisited'', {\it A\&A} {\bf 397} 147
\end{harvard}

\end{document}